**Highly flexible electromagnetic interference shielding films based on ultrathin Ni/Ag composites on paper substrates**

Xiangli Liu[c], Ziheng Ye[abc], Ling Zhang[abc], Pengdong Feng[abc], Jian Shao[abc], Mao Zhong[abc], Zheng Chen*[d], Lijie Ci[c], Peng He[b], Hongjun Ji[abc], Jun Wei[abc], Mingyu Li*[abc] and Weiwei Zhao*[abc]

[a]Flexible Printed Electronics Technology Center, Harbin Institute of Technology, Shenzhen 518055, People's Republic of China. Email: wzhao@hit.edu.cn

[b]State Key Laboratory of Advanced Welding & Joining, Harbin Institute of Technology, Shenzhen 518055, People's Republic of China. Email: myli@hit.edu.cn

[c]The School of Materials Science and Engineering, Harbin Institute of Technology, Shenzhen 518055, People's Republic of China

[d]School of Material Science and Engineering, China University of Mining and Technology, Xuzhou, Jiangsu 221116, P.R. China. Email: chenzheng1218@163.com

**Abstract**

Highly flexible electromagnetic interference (EMI) shielding material with excellent shielding performance is of great significance to practical applications in next-generation flexible devices. However, most EMI materials suffer from insufficient flexibility and complicated preparation methods. In this study, we propose a new scheme to fabricate a magnetic Ni particle/Ag matrix composite ultrathin film on a paper surface. For a ~2 μm thick film on paper, the EMI shielding effectiveness (SE) was found to be 46.2 dB at 8.1 GHz after bending 200,000 times over a radius of ~2 mm. The sheet resistance ($R_\square$) remained lower than 2.30 Ω after bending 200,000 times. Contrary to the change in $R_\square$, the EMI SE of the film generally increased as the weight ratio of Ag to Ni increased, in accordance with the principle that EMI SE is positively related with an increase in electrical conductivity. Desirable EMI shielding ability, ultrahigh flexibility, and simple processing provide this material with excellent application prospects.

*Keywords:* Electromagnetic interference shielding; Ag/Ni blend; High flexibility; Composite structure

# 1. Introduction

Flexible electronics have rapidly developed in recent years as an important direction for the future development of electronics and are of great significance to mankind. In traditional non-flexible electronic devices, solving the electromagnetic interference (EMI) problem, which interferes with delicate electronic equipment and poses environmental problems (namely long-term threats to human health), is essential in ensuring normal device operations [1-7]. EMI shielding is similarly critical in flexible electronic devices [1, 8]. For shielding materials, electrical conductivity is needed to reflect radiation by the interaction between charge carriers and electromagnetic fields [1]. In the past few years, extensive effort has been devoted to fabricating EMI shielding material with high flexibility for applications in flexible devices; strategies for effectively combing conductive fillers with flexible polymer matrix have become a common theme [1, 8-19]. One of common preparation methods is to mechanically mixing these materials, requiring a significant amount of conductive fillers to achieve desirable EMI shielding performance and can compromise flexibility. Another is to mix the conductive filler into the polymer material first, and then coat it on the surface of the flexible substrate to form a shielding material with a layered structure. However, this method not only needs to deal with the dispersion of the conductive material, how to improve the adhesion between layers is also a problem In addition to these common methods, researchers have focused on new fabrication methods to promote shielding performance and flexibility simultaneously [1, 3, 20, 21].

Metal and carbon materials are commonly used as conductive fillers for the preparation of flexible EMI shielding materials [21-24]. Among metals, silver is most commonly used because of its high conductivity, corrosion resistance, and mature production technology. In recent years, graphene (sometimes combined with metal) has been widely used in this area due to its outstanding comprehensive physical properties. For example, Zongping Chen et al. [1] prepared a graphene/polymer foam composite, where graphene sheets were fabricated by chemical vapor deposition of methane on Ni foam first, and then the Ni foam was etched away after coating a thin layer of polydimethylsiloxane (PDMS) on the graphene surface. The method provided this material (~0.8 wt% graphene, ~1 mm thick) with approximately 20.0 dB shielding effectiveness (SE) in the X band frequency, and its EMI SE decreased slightly after bending 10,000 times over a radius of 2.5 mm. Besides conductive fillers, magnetic entities are often used to improve the EMI SE of shielding materials due to the material's magnetic dipoles interacting with the radiation [2, 3, 25]. Yanhu Zhan et al. [3] obtained a flexible NR/$Fe_3O_4$@rGO composite, showing an EMI SE of 42.4 dB at 9.0 GHz (10 phr rGO, 1.8 mm thick) and an SE exhibiting only 3.5% loss after bending to a 60° angle 2000 times. These studies effectively combined flexibility with shielding properties; however, the degree of flexibility for many flexible electronic devices remains insufficient, and these shielding materials are still too thick.

In this paper, a new scheme is proposed to fabricate a highly flexible EMI shielding material with a unique composite structure by depositing and sintering Ag and Ag/Ni blends onto the paper surface. Studies have confirmed that much greater strain can be sustained by a metal film that is well bonded to substrates compared with a film that is not [26, 27]. Thus, the shielding film exhibits outstanding flexibility under highly conductive conditions owing to the contribution of the paper surface. Its EMI SE only declined 2.2 dB at 8.1 GHz with an increase of 1.20 Ω in R□ after undergoing 200,000 bending tests. The Ag/Ni blends were made from two types of Ni (Ni particles with diameters of 20-100 nm and an average diameter of 60 nm, and Ni particles with diameters of 350 nm-1.16 µm and an average diameter of 560 nm). The weight ratios of Ag to Ni were 1:2, 1:1, 2:1, 4:1, and 6:1, respectively. When the weight ratio of Ag to Ni (60 nm) reached 6:1, the Ag/Ni blend sintering film demonstrated an EMI SE larger than 46.7 dB in the 8-12 GHz frequency range and a small $R_\square$ of 0.78 Ω, while the thickness of the metal layer was only ~2 µm. Such a small thickness of the metal layer, coupled with the unique composite structure and appropriate combination of metal and paper, provides this shielding material ultrahigh flexibility and excellent shielding performance that fully satisfies application demands in many flexible electronics.

## 2. Experimental

### 2.1 Materials

Ag (50-70 nm, 99.99%), Ni (20-100 nm, 99.9%), Ni (0.34-1.12 µm, 99.8%), $AgNO_3$ (AR, 99.8%), polyvinylpyrrolidone (PVP) (Macklin, K29-32), hydrazine (Macklin, 50%), and deionized water were used in experiments.

### 2.2 Preparation of Ag/Ni blend

Ag/Ni blends were prepared by chemical reduction. First, 0.68 g of $AgNO_3$ and 0.68 g of PVP were added to 720 mL of deionized water in a beaker wrapped by in aluminum foil, and then stirred until all solvent dissolved. Then, the nano-nickel powder was added to the solvent followed by sonicating sonication of the solvent for one1 min. After adding 18 mL of hydrazine hydrate, the solvent was stirred until it became nearly transparent. The precipitate was then washed and centrifugated several times. An Ag/Ni blend was obtained after drying.

### 2.3 Preparation of EMI shielding metal/paper film

Ag/Ni blend or Ag (60 nm [representing 50-70 nm]) particles were first mixed with ethylene glycol at a weight ratio of 1:4. After sonicating and stirring for two min, the mixture was quickly coated on high-temperature resistant paper (a polyimide film coated with $TiO_2$ coating, ~100 μm thick totally) by dropper to prevent silver from depositing in liquids. Then, the paper was placed in a drying oven at 140 °C for 1 h followed by natural cooling in air. Some particles on the paper surface were not connected as a whole layer. These particles were wiped away slightly by tissue paper, and EMI shielding films were obtained.

## 2.4 Characterization

X-ray diffraction (XRD) analysis of shielding film was obtained by D/max 2500pc. The morphology and structure of Ag/Ni blend sintering films were examined using scanning electron microscopy (SEM; Hitachi S4700, 15 kV) and transmission electron microscopy (TEM; FEI Talos F200X). The thickness of the metal layer of the film was measured using a profilometer (Vecco Dektak 150). The $R_\square$ was measured with a multifunction digital four-probe tester (ST-2258C). EMI SE was measured using a vector network analyzer (Keysight, E5063A). Magnetic properties were measured by a magnetic property measurement system (Quantum Designed MPMS3).

## 3. Results and discussion

Fig. 1a shows the structure of the Ag/Ni blend sintering film in which Ag was tightly connected to the paper as an integral layer with Ni particles dispersed throughout. Our previous work has proved that the presence of $TiO_2$ in the substrate facilitates the sintering of Ag and the formation of high-strength interfaces between the Ag coating and the substrate [28]. The film sintering of the Ag/Ni blend with a continuous Ag layer was conductive, whereas the film sintering from pure Ni was not. Fig. 1b and 1c present

XRD analysis results of the substrate, pure Ag sintering films, and Ag/Ni blend sintering films. Except for the substrate peak, the film sintered from pure Ag only exhibited a diffraction peak of Ag (PDF#04-0783), while films sintered from Ag/Ni blends only demonstrated peaks of Ag (PDF#04-0783) and Ni (PDF#04-0850), demonstrating that no other substance was generated after sintering. Fig. 1d shows that the thickness of the metal layer of film, in which the weight ratio of Ag to Ni (60 nm) was 1:2, was ~2.49 µm. The total thickness of shielding film is ~100 µm, much smaller than many graphene composite shielding film [1, 3, 22]. This figure also indicates that the surface of the shielding film was not flat, confirmed by the SEM image of the film cross-section. From the SEM image, it was noted that the Ag/Ni blend has formed a metal layer tightly connected to the paper after sintering. Data were processed by ignoring the two peaks near the step and then calculating the average values at the high and low points, respectively. The height difference is regarded as the thickness of the metal layer. Fig. 1e depicts the average thicknesses of the metal layer of several shielding films; all values were between 2 µm and 2.4 µm. Each given thickness in Fig. 1e is the average of 10 values measured in different areas. The TEM and HAADF images demonstrate that magnetic Ni particles were dispersed in the continuous Ag layer as shown in Fig. 1a. This desirable structure provides this material with magnetic properties along with good electrical conductivity.

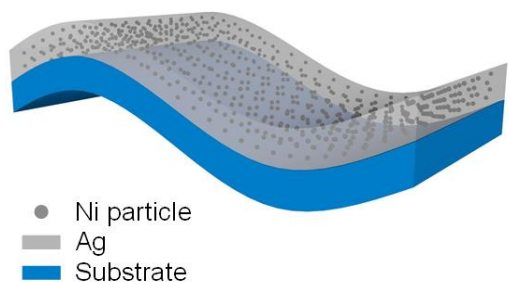
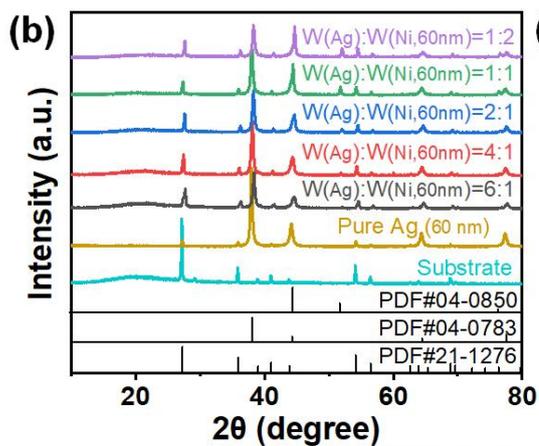
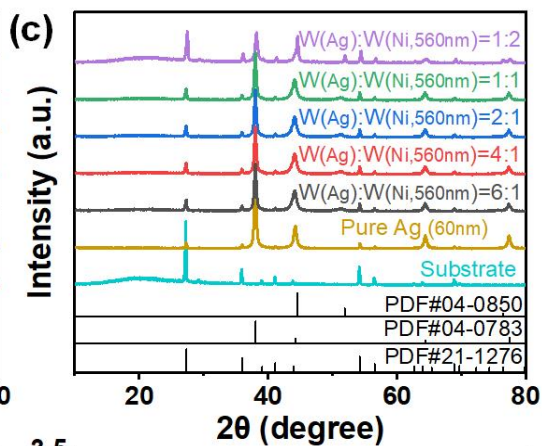
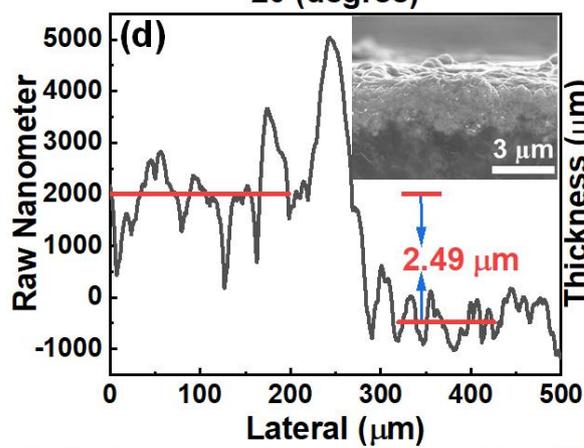
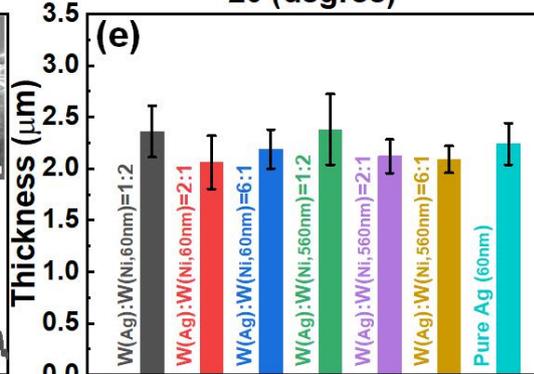
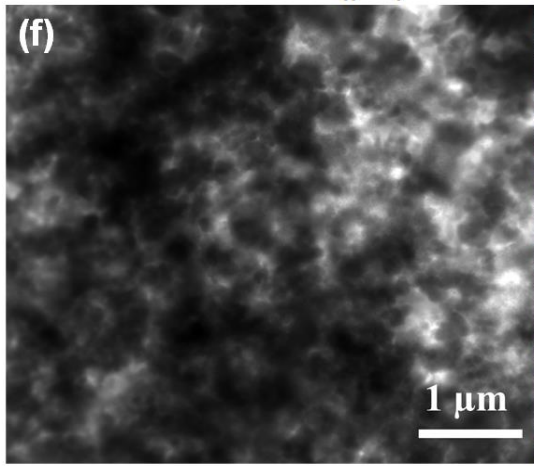
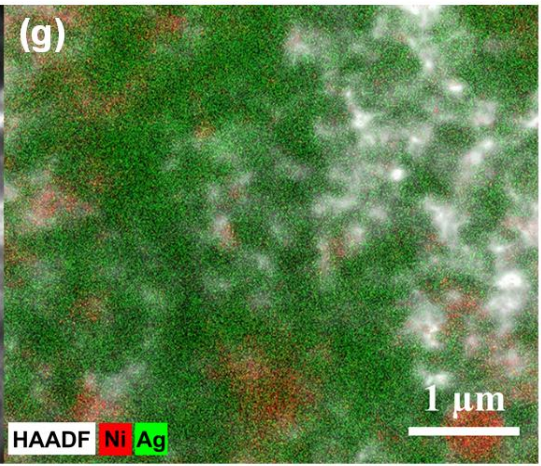

**Figure 1 a** Schematic of the Ag/Ni blend sintering film. XRD analysis of shielding films made from **b** Ni (60 nm) particles and **c** Ni (560 nm) particles. **d** Thickness of the metal layer measured by profilometer; inset is the SEM image of the film cross-section. **e** Respective average thickness of the metal layer of shielding films. **f** TEM image of the Ag/Ni blend sintering film. **g** HAADF image of the Ag/Ni blend sintering film.

Excellent flexibility under highly conductive conditions is an indispensable aspect of performance required for shielding materials in flexible electronics applications. To evaluate flexibility, several shielding films were bent to test the change in R□. For each shielding film, 0.016 g of the Ag/Ni blend or Ag particles were used, and the metal layer was slightly larger than 1 cm× 1 cm. All R□ were measured in the bending area. Fig. 2a and 2b show the change in R□ for the shielding films, indicating that all films exhibited outstanding flexibility. The R□ of these shielding films rose slightly during 70,000 bending tests, and the variation tendency became stronger thereafter. The R□ also declined with an increase in Ag content and approximate reduction in Ni content. Nevertheless, the reduction in R□ was inconspicuous as the content of Ag increased while the weight ratio of Ag to Ni reached a certain value (1:1 for films made from Ni [60 nm] particles or 2:1 for films made from Ni [560 nm] particles); thus, no considerable effects were caused by the presence of a relatively small amount of Ni on the sintering of the Ag/Ni blend.

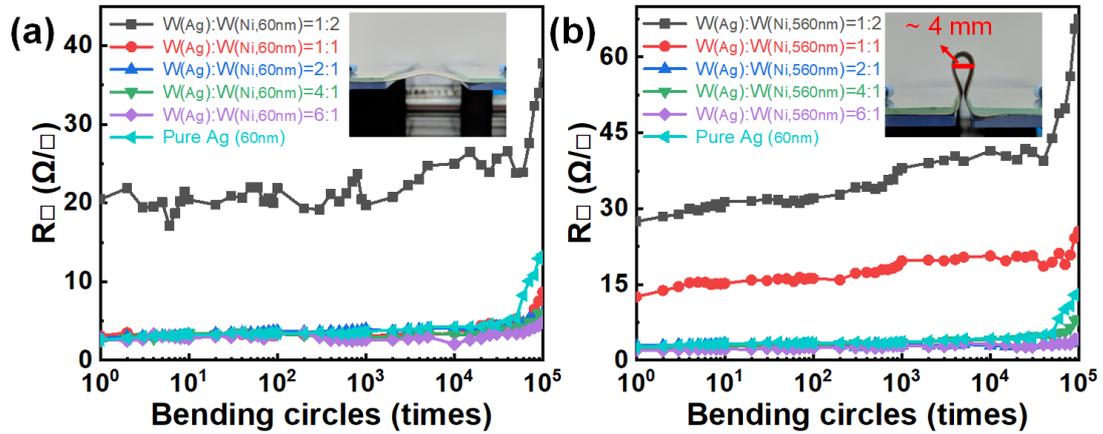

**Figure 2** R□ changes in the bending area of shielding films made from **a** Ni (60 nm) particles and **b** Ni (560 nm) particles after bending. The left inset shows the start state and end state of a bending circle; the right inset shows the minimum radius (~2 mm) of a bending circle.

Owing to the high electrical conductivity of Ag, small thickness of the metal layer, and appropriate combination of the metal and substrate, the R□ of the film made from pure Ag increased gradually to only ~5.40 Ω after bending 50,000 times and to ~13.00 Ω after bending 100,000 times. Notably, when the weight ratios of Ag to Ni (60 nm) were 1:1, 2:1, 4:1, and 6:1, the Ag/Ni blend sintering films demonstrated similar R□ of less than roughly 5.00 Ω after bending 60,000 times and less than 9.00 Ω after bending 100,000 times. Given the unique composite structure and flexible paper, these Ag/Ni blend sintering films were highly flexible and conductive. However, when the weight ratio of Ag to Ni (60 nm) reached 1:2, the sintering of the Ag/Ni blend was greatly affected by the high amount of Ni. Thus, the R□ of the corresponding shielding film was largest but no higher than 28.00 Ω after bending 70,000 times and 38.00 Ω after bending 100,000 times, among all shielding films shown in Fig. 2a.

Similar to the electrical performance of shielding films made from Ni (60 nm) particles, the Ag/Ni blend sintering films at Ag:Ni (560 nm) weight ratios of 2:1, 4:1, and 6:1 showed low R▫, less than 6.00 Ω, after bending 70,000 times, and 9.00 Ω after bending 100,000 times. Yet when the weight ratios of Ag to Ni (560 nm) were 1:2 and 1:1, the corresponding films had higher R▫ compared with other films made from Ni (560 nm) particles. The R▫ of the 1:2 ratio was less than 50.00 Ω after bending 70,000 times and 68.00 Ω after bending 100,000 times, whereas that of the 1:1 ratio was less than 19.00 Ω after bending 70,000 times and 26.00 Ω after bending 100,000 times.

Using shielding material to block electromagnetic waves is a common and effective way to solve the EMI pollution problem. Generally, reflection ($SE_R$), absorption ($SE_A$), multiple reflection ($SE_M$), and transmission are generated while electromagnetic waves encounter shielding material; the sum of the first three factors is calculated as the total EMI SE ($SE_T$). The $SE_A$, $SE_R$, and $SE_T$ can be calculated by combining the power coefficients of reflection (R), absorption (A), and transmission (T). The corresponding expressions are as follows:

$$SE_T = -10\log T = SE_R + SE_A + SE_M \tag{1}$$

$$SE_R = -10\log_{10}(1-R) \tag{2}$$

$$SE_A = -10\log_{10}[T/(1-R)] \tag{3}$$

The $SE_M$ is generally ignored when $SE_T$ is larger than 15.0 dB.

Fig. 3a and 3b present the EMI SE and R▫ of several shielding films measured in the 8-12 GHz frequency range. For each shielding film, 0.144 g of Ag or the Ag/Ni blend were used; the metal layer was 3 cm× 3 cm. The sample holder of the vector network analyzer with shielding film is shown in Fig.. S1. Electrical conductivity of a shielding material is highly significant to EMI SE. The substrate barely shielded the electromagnetic wave due to its non-conductivity, resulting in an EMI SE of approximately 0 dB. Conversely, the pure Ag sintering film was highly conductive with an EMI SE greater than 36.0 dB within 8-12 GHz (the maximum was 43.0 dB at 8 GHz), implying it can block more than 99.97% of electromagnetic waves. The R□ of some pure Ag sintering films was smaller than 1.84 Ω; The film with an R□ of 1.84 Ω was selected for comparison with other Ag/Ni blend sintering films. When the weight ratio of Ag to Ni (60 nm) was 2:1, the EMI SE of the Ag/Ni blend sintering film was only 1.0 dB smaller than that of the pure Ag sintering film, on average, within 8-12 GHz; the R□ of the Ag/Ni blend sintering film was 0.82 Ω larger than that of the pure Ag film. However, when the weight ratio of Ag to Ni (60 nm) was 6:1, a difference in R□ of only 0.02 Ω emerged between the pure Ag sintering film and the Ag/Ni blend sintering film; the EMI SE of the latter was 4.7 dB larger on average than that of the former at 8-12 GHz, proving that the presence of magnetic Ni particles can enhance EMI shielding performance of the films. The 53.5 dB high EMI SE of the Ag/Ni blend sintering film is much larger than most of flexible metal-based sponge and carbon-based (carbon nanotubes, graphene) shielding composites, although its thickness is much smaller. An Examples are the silver nanowire wrapped carbon core-shell hybrid sponge (Ag@C) with 37.9 dB EMI SE at 1 mm, and the graphene/PDMS foam composite with 20 dB EMI SE at 1 mm [1, 29]. Generally, it is difficult to form a connected conductive network in insulative polymer matrices with low conductive fillers. Thus, a high filler content and large thickness are necessary to achieve desirable shielding ability. The examples mentioned are well equipped with shielding performance compared with other same-kind material but still poor in shielding ability compared with the experimental shielding film with a continuous silver layer and magnetic Ni particles. In addition, fabricating these two materials is challenging; the former needs chemical vapor deposition and the latter needs annealing at 1000 °C in an argon atmosphere.

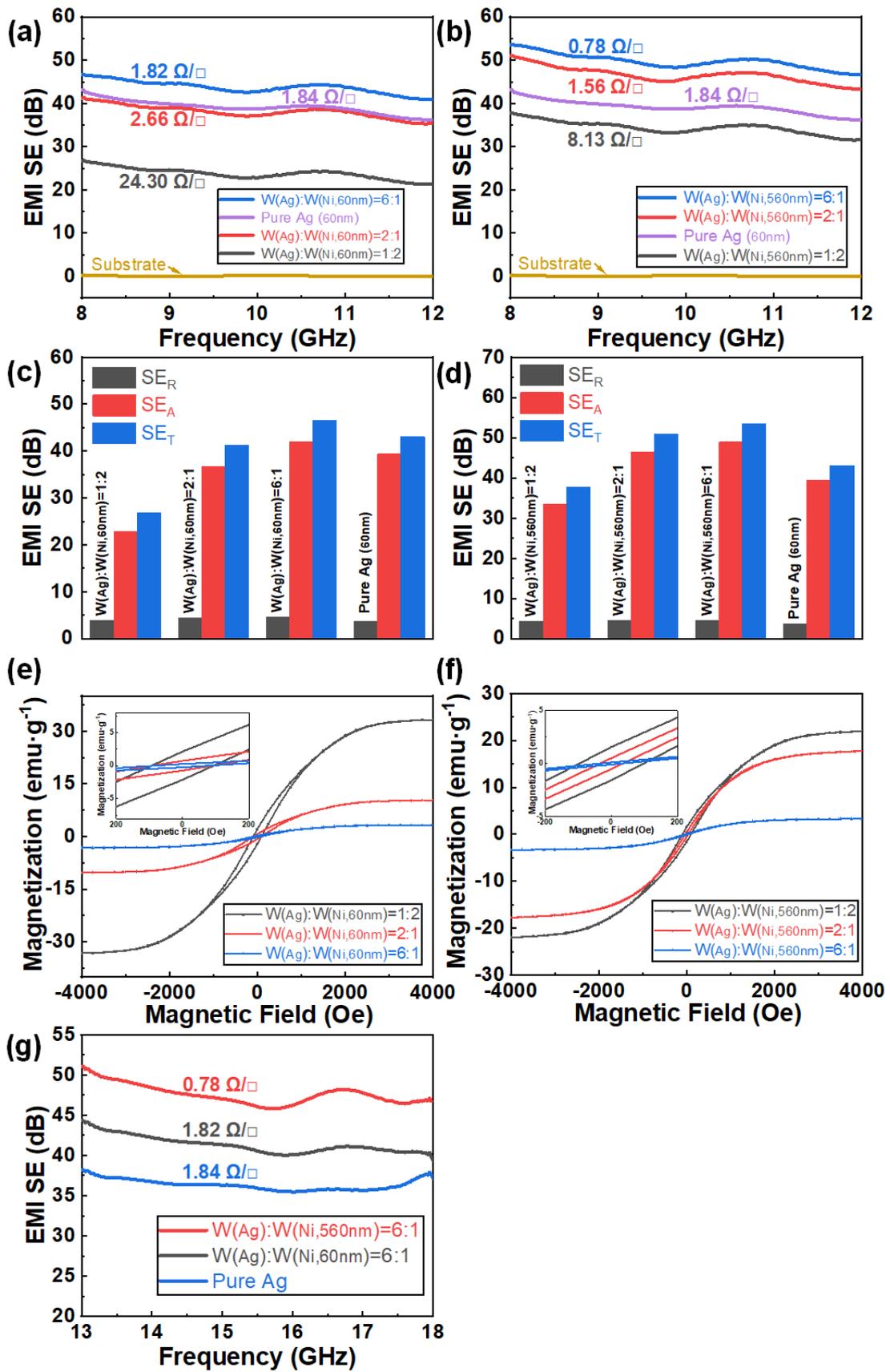

**Figure 3** EMI SE of shielding films made from **a** Ni (60 nm) particles and **b** Ni (560 nm) particles measured in the 8-12 GHz frequency range. Comparison of $SE_T$, $SE_R$, and $SE_A$ of films made from **c** Ni (60 nm) particles and **d** Ni (560 nm) particles at 8 GHz. Magnetic properties of metal layer of films made from **e** Ni (60 nm) particles and **f** Ni (560 nm) particles. Inset images are amplified views of magnetization vs. magnetic field. **g** EMI SE of shielding films measured in the 13-18 GHz frequency range.

Combining Fig. 3a and 3b, the film made from Ni (560 nm) particles showed lower $R_\square$ and higher EMI SE compared with that made from Ni (60 nm) particles when the two films shared the same weight ratio of Ag to Ni. For example, at an Ag to Ni weight ratio of 6:1, the EMI SE of the film made from Ni (560 nm) particles was 53.5 dB at 8 GHz, whereas that of the film made from Ni (60 nm) particles was 46.6 dB at 8 GHz. As mentioned earlier, the $R_\square$ of Ag/Ni blend sintering films increased as the weight ratio of Ni to Ag increased. Unsurprisingly, the shielding film with a 2:1 weight ratio of Ag to Ni (560 nm) had a $R_\square$ of 1.56 Ω and an EMI SE of 50.9 dB at 8 GHz, whereas the film with a 1:2 weight ratio of Ag to Ni (560 nm) exhibited 8.13 Ω and 37.7 dB at 8 GHz, respectively.

The thickness of a shielding material significantly influences its EMI SE (Note 1 in Appendix). To determine the shielding performance of a material more realistically, it is reasonable to use the SSE/t as an evaluation criterion, wherein the specific EMI shielding effectiveness (SSE) is divided by the thickness of the shielding material to yield a normalized value. Surprisingly, when the weight ratio of Ag to Ni (560 nm) was 6:1, the metal layer of the Ag/Ni blend sintering film had a high SSE/t of 65,224 dB cm$^2$ g$^{-1}$, much larger than that of copper foil (7,812 dB cm$^2$ g$^{-1}$), aluminum foil (30,555 dB cm$^2$ g$^{-1}$), and a Ti$_3$C$_2$Tx MXene film (30,830 dB cm$^2$ g$^{-1}$) [2, 29].

Shielding of an electromagnetic wave primarily originates from reflection and absorption mechanisms; reflection results from a mismatch between the absorber and air, whereas absorption occurs from ohmic loss, polarization loss, and magnetic loss [3, 25]. To further analyze the shielding mechanism in EMI SE results, $SE_R$ and $SE_A$ were calculated. Fig. 3c and 3d show the $SE_R$, $SE_A$, and $SE_T$ of several shielding films. All films exhibited similar $SE_R$ values but different $SE_A$ and $SE_T$ values at 8 GHz. Clearly, $SE_A$ and $SE_T$ each increased with a decline in R□, while $SE_A$ remained generally stable. Owing to the contribution of magnetic Ni particles, the Ag/Ni blend sintering film with a weight ratio of Ag to Ni (60 nm) of 2:1 showed a larger $SE_A$ than the pure Ag sintering film; however, the R□ of these two films was approximately the same. For any shielding film in Fig. 3c or 3d, $SE_A$ was much larger than $SE_R$, implying that absorption contributed more to EMI SE compared to reflection. Considering the damage to the reflected portion of electromagnetic waves on the surroundings, this absorption-dominant EMI shielding material is preferred in areas that require EMI shielding and generate electromagnetic radiation [1].

Fig. 3e and 3f illustrate magnetic properties of the shielding films. Similar to the change in R□, the remnant magnetization (Mr) and saturation magnetization (Ms) of metal layers made from the same Ni particles increased with an increase in Ni content. For instance, the R□, Mr, and Ms of the metal layer with a weight ratio of Ag to Ni (60 nm) of 6:1 were 1.82 Ω, 0.25 emu g$^{-1}$, and 3.01 emu g$^{-1}$, respectively, whereas those of the metal layer with a weight ratio of Ag to Ni (60 nm) of 2:1 were 2.66 Ω, 0.73 emu g$^{-1}$, and 10.25 emu g$^{-1}$. Interestingly, a coercivity (Hc) as high as 94.0 Oe was found in films made from Ni (60 nm) particles, but the Hc of films made from Ni (560 nm) particles with Ag:Ni weight ratios of 1:2, 2:1, and 6:1 were 97.4 Oe, 34.1 Oe, and 34.1 Oe, respectively. This difference is primarily due to different Ni particle sizes [30]. Although distinct magnetic properties were identified in the films made from Ni (60 and 560 nm), the differences in R□ primarily distinguished the EMI SE of these films.

Generally, the problem of EMI SE declining in the high-frequency band due to stronger penetration of high-frequency electromagnetic waves poses challenges to many shielding materials [31]. Similar results also occurred in these experiments. To further evaluate changes in the shielding effect at different frequencies, our shielding films were measured at 8-12 and 13-18 GHz frequency ranges. Fig. G displays the EMI SE of shielding films in the 13-18 GHz frequency range. For the film with an Ag:Ni (560 nm) weight ratio of 6:1, the EMI SE was larger than 45.8 dB in the 13-18 GHz frequency range. With the aid of magnetic Ni (60 nm) particles, the EMI SE of the Ag/Ni blend sintering film with a weight ratio of Ag to Ni of 6:1 was ~5.0 dB higher than that of the pure Ag sintering film. The shielding performance of these films generally decreased but remained satisfactory for applications in many shielding fields as the frequency increased.

Stable shielding performance under mechanical deformation is urgently needed for shielding material applications in flexible electronics. Fig. 4a depicts changes in R□ after repeated bending. The metal layer of each film measured ~3 cm×3 cm. The initial R□ measured in the bending area was lower than the average R□ of the shielding film. As a main factor affecting EMI SE, the R□ of the pure Ag sintering film increased by 2.96 Ω, whereas that of the Ag/Ni blend sintering film only increased by 1.20 Ω. Fig. 4b shows the change in EMI SE after repeated bending. Encouragingly, after bending 200,000 times, the EMI SE of the pure Ag sintering film and Ag/Ni blend sintering film only decreased by 11.7% and 4.5% at 8.1 GHz, respectively. Owing to the reinforcement of Ni particles, the Ag/Ni blend sintering film with a composite structure exhibited more stable shielding and electrical performance than the pure Ag sintering film [32]. These films exhibit much better flexibility than popular flexible graphene composites. The representative composites, as mentioned before, are a graphene/polymer foam composite and a NR/Fe$_3$O$_4$@rGO composite [1, 3]. EMI SE of the former only showed a slight decrease after 10,000 times bending to a radius of 2.5 mm while the latter could withstand 2000 times bending at an angle of 60°. However, higher degree of flexibility is needed for practical applications in flexible electronics. High loading of conductive fillers, weak connection of conductive networks, large thickness of shielding material, and poor binding between fillers and matrix still challenge many traditional flexible shielding materials [1, 3, 33]. For the studied shielding film, a desirable EMI SE (above 46.7 dB within 8-12 GHz), low density (1.39 cm$^3$ g$^{-1}$), small low thickness (thickness of the metal layer was ~2 μm and that of the total film was ~100 μm, although the paper can be made much thinner than 100 μm), excellent flexibility (capable of undergoing 200,000 bending tests), and simple processing (sintering Ag/Ni blend on papers at 140 °C) of this shielding film fully satisfy the needs of commercial applications.

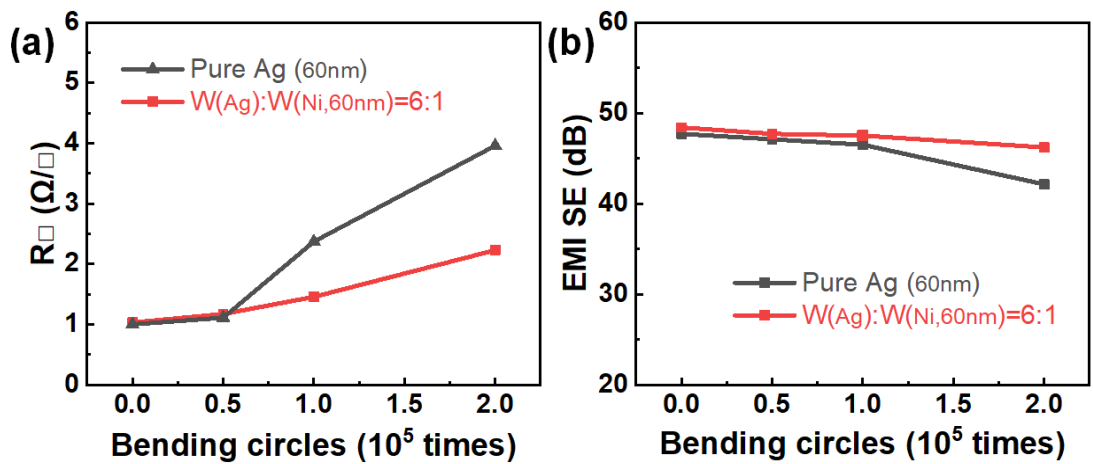

**Figure 4 a** R☐ change in bending area of shielding films after bending. **b** EMI SE change of shielding films at 8.1 GHz after bending (bending radius: ~2 mm).

**4. Conclusions**

In summary, we developed a flexible EMI shielding film with a unique composite structure using a simple fabrication method that directly sinters Ag/Ni blend onto paper. The ~2-µm-thick metal layer endowed the shielding material with a high EMI SE of more than 46.7 dB within the 8-12 GHz frequency range and a small R☐ of 0.78 Ω. As the content of Ag declined, R☐ increased, leading to a reduction in EMI SE. After bending to a radius of ~2 mm 200,000 times, the shielding film with an Ag:Ni (60 nm) weight ratio of 6:1 only exhibited a 4.5% loss in EMI SE and an increase of 1.20 Ω in R☐. Such stable shielding performance under mechanical deformation combined with simple processing and small thickness provide this material strong application potential in next-generation flexible electronics.

**Conflicts of interest**

There are no conflicts to declare.

**Acknowledgements**

We thank Guojian Cao and Guohua Fan for the TEM measurements and Sixia Hu in Southern University of Science and Technology for the measurements on magnetic properties. This work was supported by Shenzhen Science and Technology Program (Grant No. KQTD20170809110344233, Grant No. JCYJ20170811160129498) and Bureau of Industry and Information Technology of Shenzhen through the Graphene Manufacturing Innovation Center (201901161514). Z.C. acknowledges the Natural

Science Foundation of China (No.51771226). X.L. acknowledges the Natural Science Foundation of China (No. 11672090).

**Author contribution**

W.Z. designed the experiments. X.L., Z.Y. L.Z. made the paper-based devices with the help of W.Z., P.H., J.W.. X.L., Z.Y., P.F., J.S., M.Z. did the measurements on flexibility properties with the help of W.Z., H.J. and M.L.. Z.Y. performed measurements of the electromagnetic interference shielding with the help of L.C.. Z.C. made the Ni nano particles. Z.Y., M.L. and W.Z. wrote the Manuscript. All the authors participated in the analysis of the data and the preparation of the final manuscript.